\documentstyle[12pt]{article}
\newfont{\ffont}{msym10}                        
\newcommand{\beq}{\begin{equation}}             
\newcommand{\eeq}{\end{equation}}               
\newcommand{\bqry}{\begin{eqnarray}}            
\newcommand{\eqry}{\end{eqnarray}}              
\newcommand{\bqryn}{\begin{eqnarray*}}          
\newcommand{\eqryn}{\end{eqnarray*}}            
\newcommand{\NL}{\nonumber \\}                  
\newcommand{\preprint}[1]{\begin{table}[t]      
            \begin{flushright}                  
            \begin{large}{#1}\end{large}        
            \end{flushright}                    
            \end{table}}                        
\newcommand{\PD}[2]                             
    {\frac{\partial^{#2}}{\partial #1^{#2}}}    
\begin{document}
\preprint{TAUP-2081-93 \\ }
\title{Galilean Limit of
Equilibrium Relativistic Mass Distribution}
\author{\\ L. Burakovsky\thanks {Bitnet:BURAKOV@TAUNIVM.} \
and L.P. Horwitz\thanks
  {Bitnet:HORWITZ@TAUNIVM. Also at Department of Physics, Bar-Ilan
University, Ramat-Gan, Israel  } \\ \ }
\date{School of Physics and Astronomy \\ Raymond and Beverly Sackler
Faculty of Exact Sciences \\ Tel-Aviv University,
Tel-Aviv 69978, ISRAEL}
\maketitle
\begin{abstract}
The low-temperature form of the equilibrium relativistic mass
distribution is subject to the Galilean limit by taking $c\rightarrow
\infty .$ In this limit the relativistic Maxwell-Boltzmann distribution
passes to the usual nonrelativistic form and the
Dulong-Petit law is recovered.
\end{abstract}
\bigskip
{\it Key words:} special relativity, relativistic Maxwell-Boltzmann,
mass distribution, Galilean limit.

PACS: 03.30.+p, 05.20.Gg, 05.30.Ch, 98.20.--d
\bigskip
\section{Introduction}

In a previous paper \cite{BH} we studied an equilibrium relativistic
ensemble, described by an equilibrium relativistic Maxwell-Boltzmann
distribution with variable mass. For such a system a
well-defined mass distribution was found,
consistent in low-temperature limit with the
one obtained by Hakim \cite{Hak} from the well-known J\"{u}ttner-Synge
distribution \cite{Syn} of an on-mass-shell relativistic kinetic theory.
Calculations of the average values of mass and energy gave in the
low-temperature limit a correction of the order of $10\%$ to the
Dulong-Petit law.

In the present paper we consider the Galilean limit of low-temperature
form of the equilibrium relativistic mass distribution. We show that no
correction to the Dulong-Petit law appears in this limit of the theory.

\section{Preliminary remarks}

In a previous paper, having begun with the low-temperature form of the
relativistic Maxwell-Boltzmann distribution [1,(47)] (we use the metric
$g^{\mu \nu }=(-,+,+,+),\;\;q\equiv q^\mu ,\;\;p\equiv p^\mu ,$
and take $\hbar =c=1$ unless otherwise specified)
\beq
f(p,q)=C(q)e^{-Am_c^2}e^{2Ap_\mu p_c^\mu },
\eeq
which coincides with the
J\"{u}ttner-Synge distribution adopted for an on-mass-shell relativistic
kinetic
theory, we obtained the following low-temperature form of the equilibrium
relativistic mass distribution [1,(48)]:
\beq
f(m)=\frac{(2Am_c)^3}{2}m^2K_1(2Am_cm),
\eeq
where $m_c=\sqrt{-p_c^\mu p_{c\mu }}$ and
$K_1$ is the Bessel function of the third kind (imaginary
argument),
\beq
K_{\nu }(z)=\frac{\pi i}{2}e^{\pi i\nu /2}H_{\nu }^{(1)}(iz).
\eeq
In this formula $2Am_c$ corresponds to $1/k_BT$ [1,(13)].

The distributions (1),(2) gives the following values of the average
four-momentum and mass: $$\langle p^\mu \rangle =4\frac{p_c^\mu }{m_c}k_B
T,\;\;\langle m\rangle =\frac{3\pi }{4}k_BT.$$ In the local rest frame we
have
\beq
\langle E\rangle=4k_BT
\eeq
and, consequently,
\beq
\langle E\rangle -\langle m\rangle =\gamma \frac {3}{2}k_BT,
\eeq
where $\gamma =\frac{16-3\pi }{6}\approx 1.1$ represents a relativistic
correction to the Dulong-Petit law.

This result follows directly from equilibrium thermodynamics without
imposing the geometrical restriction of the precise Galilean group to an
infinitely sharp mass shell. The reason for appearance of such a
correction is determined by the fact that the difference $E-m,$ even in
the low-temperature limit, does not correspond to the expression for
nonrelativistic energy $\frac{{\bf p}^2}{2m}.$ Therefore, the result
that we have found can not be considered as a nonrelativistic limit;
it is actually a relativistic low-temperature limit alone.

As is well known \cite{LevyL}, the structure of the Galilean group,
the symmetry of nonrelativistic system, implies that the mass of a
particle must be a constant intrinsic property.

In this paper we shall consider the Galilean limit of equilibrium
relativistic ensemble which has been treated in the series of papers
[1],\cite{HorShaSch}, by taking $c\rightarrow \infty $
\cite{HSP},\cite{HR}. We shall see that in the Galilean limit the
difference $E-m$ approaches the nonrelativistic expression
$\frac{{\bf p}^2}{2M},$ where Galilean mass $M$ coincides with the
particle's intrinsic parameter. This variable $\frac{{\bf p}^2}{2M}$
turns out to be distributed over ensemble with the usual
nonrelativistic Maxwell-Boltzmann distribution, due to the fact that the
relativistic relation
between the energy $E$ and the mass $m$ $E^2=m^2+{\bf p}^2$
transforms in the Galilean limit to $E=m+\frac{{\bf p}^2}{2M},$
giving rise to the Maxwell-Boltzmann distribution of the latter. The
first moment of this distribution $\langle \frac{{\bf p}^2}{2M}
\rangle $ (which coincides with $\langle E-m\rangle ),$ takes the value
$\frac{3}{2}k_BT,$ in agreement with the Dulong-Petit law.

We recognize, however, that the applicability of the Galilean group is an
idealization of a world which seems to be more correctly described by the
Poincar\'{e} group, and the Galilean limit is just a reasonable
approximation for the relativistic relation $E^2=m^2+{\bf p}^2.$

\section{Galilean limit of a free relativistic $N$-particle system}
We consider a system of $N$ particles in the framework of a manifestly
covariant mechanics \cite{rqm}, both for the classical theory and for the
corresponding relativistic quantum theory. For the classical case, the
dynamical evolution of such a system is governed by the equations of
motion that are of the form of Hamilton equations for the motion of $N$
$events$ which generate the space-time trajectories (particle world
lines). These events are considered as the fundamental dynamical objects
of the theory; they are characterized by their positions $q^\mu =(ct,{\bf
r})$ and energy-momenta $p^\mu =(E/c,{\bf p})$ in an $8N$-dimentional
phase space. The motion is parametrized by an invariant parameter $\tau $
[8], called the ``historical time''. The collection of events (called
``concatenation'' \cite{conc}) along each world line corresponds to a
$particle,$ and hence the evolution of the state of the $N$-event system
describes, $a\;posteriori,$ the history in space and time of an
$N$-particle system.

For the quantum case the dynamical evolution is governed by a generalized
Schr\"{o}dinger equation for the wave function $\psi _\tau (q_1,q_2,
\ldots ,q_N)\in L^2(R^{4N}),$ the Hilbert space of square integrable
functions with measure $dq_1dq_2\cdots dq_N\equiv d^{4N}q,$ describing
the distribution of events and representing the probability amplitudes
for finding events at space-time points $(q_1^\mu ,q_2^\mu ,\ldots ,q_N^
\mu )$ at any instant $\tau :$
$$i\hbar \frac{\partial \psi _\tau (q_1,\ldots ,q_N)}{\partial \tau }=
K\psi _\tau (q_1,\ldots ,q_N),$$ where $K$ is the dynamical evolution
operator (generalized Hamiltonian), of the same form
for both classical and quantum cases.

We shall consider here
a many-particle system, within the framework of the relativistic
generalization of the usual nonrelativistic Boltzmann theory [5].

To study the nonrelativistic limit of a dilute gas of events, it is
sufficient to treat the simplest case of a system of $N$ free particles
with the Hamiltonian
\beq
K_0=\sum _{i=1}^N\frac{p_{i\mu }p_i^\mu }{2M_i},
\eeq
where $M_i$ are positive parameters, the given intrinsic properties of
the particles, having the dimension of mass.

The Hamilton equations
$$\frac{dq_i^\mu }{d\tau }=\frac{\partial K}{\partial p_
{i\mu }},\;\;\;\frac{dp_i^\mu }{d\tau }=-
\frac{\partial K}{\partial q_{i\mu }},\;\;\;i=1,2,\ldots ,N$$
yield, in this case,
$$\frac{dq_i^\mu }{d\tau }=\frac{p_i^\mu }{M_i},\;\;\;
\frac{dp_i^\mu }{d\tau }=0,\;\;\;i=1,2,\ldots ,N.$$

The evolution of the wave function is described by the equation
\beq
i\hbar \frac{\partial \psi _\tau (q_1,\ldots ,q_N)}{\partial \tau }=K_0
\psi _\tau (q_1,\ldots ,q_N).
\eeq
The wave function can be expressed as a Fourier transform
\begin{eqnarray}
\psi _\tau (q_1,\ldots ,q_N) & = &
\frac{1}{(2\pi \hbar )^{4N}}\int d^4p_1\cdots d^4p_Ne^{\frac{i}{\hbar }
\sum _{k=1}^Np_k^\mu q_{k\mu }}\psi _\tau (p_1,\ldots ,p_N) \nonumber \\
  & = &
\frac{1}{(2\pi \hbar )^{4N}}\int d^4p_1\cdots d^4p_Ne^{\frac{i}{\hbar }
\sum _{k=1}^Np_k^\mu q_{k\mu }}e^{-\frac{i}{\hbar }K_0\tau }
\psi _0(p_1,\ldots ,p_N) \nonumber \\   & = &
\frac{1}{(2\pi \hbar )^{4N}}\int d^4p_1\cdots d^4p_Ne^{\frac{i}{\hbar }
\sum _{k=1}^N(p_k^\mu q_{k\mu }-\frac{p_k^\mu p_{k\mu }}{2M_k}\tau )}
\psi _0(p_1,\ldots ,p_N).
\end{eqnarray}
If this wave function is to be associated with particles, the function
$\psi _\tau (p_1,\ldots ,p_N)=e^{-i/\hbar K_0\tau }\psi _0(p_1,\ldots ,
p_N)$ must have support in momentum space in a region which is in the
neighborhood of definite masses (as pointed out in [7], these
considerations are valid also in the presence of interaction, if it is
not too strong). In the nonrelativistic limit this support should
approach the corresponding
definite mass shells, consistent with a representation of the
Galilean group.

We shall, therefore, require that the quantities
\beq
\epsilon _i=E_i-M_ic^2,\;\;\;i=1,\ldots ,N,
\eeq
constructed of variables occuring in the integrand of (8), i.e., in the
support of $\psi _\tau (p_1,\ldots ,p_N),$ be finite as $c\rightarrow
\infty $ (compared to all other velocities) for the states with finite
momenta.

We shall see that it is sufficient that the support of $\psi _\tau (p_1,
\ldots ,p_N)$ contract such that the variables
\beq
\eta _i=c^2(m_i-M_i),\;\;\;i=1,\ldots ,N
\eeq
may take any value, however, finite, as $c\rightarrow \infty ;$
or, equivalently,
\beq
m_i=M_i(1+O(\frac{1}{c^2})).
\eeq
(The situation is quite similar to one in relativistic classical
statistical mechanics, when this freedom permits one to obtain the
Galilean microcanonical ensemble [6].)

Indeed, in this case one can show that the values $E_i-m_ic^2$ are
equal to
\beq
E_i-m_ic^2=\frac{{\bf p}_i^2}{2M_i}+O(\frac{1}{c^2}),\;\;\;i=1,
\ldots ,N
\eeq
and approach nonrelativistic kinetic energies of particles with the
Galilean masses $M_i$ as $c\rightarrow \infty :$
\bqry
E_i-m_ic^2 & = & \sqrt{(m_ic^2)^2+{\bf p}_i^2c^2}-m_ic^2 \NL      & = &
\sqrt{(\eta _i+M_ic^2)^2+{\bf p}_i^2c^2}-m_ic^2  \NL         & = &
M_ic^2\sqrt{1+\frac{2\eta _i}{M_ic^2}+\frac{\eta _i^2}{M_i^{2}c^4}+\frac
{{\bf p}_i^2}{M_i^{2}c^2}}-m_ic^2 \NL  & = & M_ic^2(1+\frac{\eta _i}
{M_ic^2}+\frac{{\bf p}_i^2}{2M_i^{2}c^2}+O(\frac{1}{c^4}))-m_ic^2 \NL
& = & (M_ic^2+\eta _i)+\frac{{\bf p}_i^2}{2M_i}-m_ic^2+O(\frac{1}{c
^2}) \NL      & = & \frac{{\bf p}_i^2}{2M_i}+O(\frac{1}{c^2}).
\eqry
Consequently, the quantities
\beq
\epsilon _i=E_i-M_ic^2=E_i-m_ic^2+(m_i-M_i)c^2=
\frac{{\bf p}_i^2}{2M_i}+\eta _i+O(\frac{1}{c^2}),\;\;\;
i=1,\ldots ,N,
\eeq
are finite as $c\rightarrow \infty ,$ as was required from the very
beginning.

Now we turn to investigate the behavior of the wave function $\psi _\tau
(q_1,\ldots ,q_N)$ in the Galilean limit:
\bqry
\psi _\tau (q_1,\ldots ,q_N) & = &
\frac{1}{(2\pi \hbar )^{4N}}\int d^4p_1\cdots d^4p_Ne^{\frac{i}{\hbar }
\sum _{k=1}^N({\bf p}_k{\bf r}_k-E_kt_k)}
e^{\frac{i}{\hbar }\sum _{k=1}^N\frac{m_k^2c^2}{2M_k}\tau }
\psi _0(p_1,\ldots ,p_N) \nonumber \\           & = &
\frac{1}{(2\pi \hbar )^{4N}}\int \frac{dE_1}{c}\cdots \frac{dE_N}{c}
d^3{\bf p}_1\cdots d^3{\bf p}_Ne^{\frac{i}{\hbar }
\sum _{k=1}^N[{\bf p}_k{\bf r}_k-(\frac{{\bf p}_k^2}{2M_k}+M_k
c^2+\eta _k)t_k]} \NL       &      &    \times
e^{\frac{i}{\hbar }\sum _{k=1}^N(\frac{M_kc^2}{2}+\eta _k)\tau }
\psi _0(p_1,\ldots ,p_N) \nonumber \\            & = &
\frac{1}{(2\pi \hbar )^{4N}}\int \frac{d\eta _1}{c}\cdots \frac{d\eta _N}
{c}d^3{\bf p}_1\cdots d^3{\bf p}_Ne^{\frac{i}{\hbar }
\sum _{k=1}^N({\bf p}_k{\bf r}_k-\frac{{\bf p}_k^2}{2M_k}t_k)}
e^{-\frac{i}{\hbar }\sum _{k=1}^NM_kc^2t_k} \NL    &    &   \times
e^{\frac{i}{\hbar }\sum _{k=1}^N\frac{M_kc^2}{2}\tau }e^{\frac{i}{\hbar }
\sum _{k=1}^N(\tau -t_k)\eta _k}\psi _0(p_1,\ldots ,p_N).
\eqry
Although the support of $\psi _0(p_1,\ldots ,p_N)$ is bounded in the
$\eta _k$'s as $c\rightarrow \infty ,$ the integrals over the $\eta _k$'s
can approximately yield factors of $\delta (t_k-\tau ),$ as remarked
in [7]. Consider the case for which $\psi _0(p_1,\ldots ,p_N)$ is
independent of $\eta _k,$ for $k=1,\ldots ,N,$
in $-\triangle _k\leq \eta _k\leq \triangle _k$ and is
zero outside this region; then the wave function (15) is
proportional to the product
\bqry
\prod _{k=1}^N\int _{-\triangle }^\triangle d\eta _ke^{\frac{i}{\hbar }
\eta _k(\tau -t_k)} & = &
\prod _{k=1}^N\int _{-\triangle /\hbar }^{\triangle /\hbar }\hbar
d\eta _k^{'}e^{i\eta _k^{'}(\tau -t_k)} \NL   & = &
(2\pi \hbar )^N\prod _{k=1}^N\delta  _{\triangle /\hbar }(\tau -t_k),
\eqry
where $$\triangle =\min (\triangle _1,\triangle _2,\ldots ,\triangle _N)
$$ and $$\delta _{\triangle /\hbar }(\tau -t_k)\rightarrow \delta (\tau
-t_k),$$ if\footnote{$\triangle \rightarrow \infty $ may also satisfy
this condition; in the present paper we shall use the fact that
$\triangle $ can take any infinitesimal value but not zero.}
$\hbar \rightarrow 0$ (it is clear now that $\triangle \rightarrow 0$
precisely would be an unsuitable condition for the nonrelativistic
limit). The dispersion of $t_k$ around $\tau ,$ bounded by $$\mid t_k-
\tau \mid \leq \hbar /\triangle _k\leq \hbar /
\triangle $$ is therefore a purely quantum effect (it does not depend on
$c$ and vanishes with $\hbar \rightarrow 0),$
emerging asymptotically from a relativistic quantum theory in the
Galilean limit, as emphasised in [7].

Thus we will have
\beq
\int \frac{d\eta _1}{c}\cdots \frac{d\eta _N}{c}e^{\frac{i}{\hbar }\sum
_{k=1}^N(\tau -t_k)\eta _k}\cong
(2\pi \hbar )^N\prod _{k=1}^N\delta  (c\tau -ct_k),
\eeq
which means that the times associated with all of the particles become
synchronized in the Galilean limit:
\beq
ct_1=ct_2=\ldots =ct_N=ct=c\tau .
\eeq
This result also can be obtained from the canonical equations of motion:

we have with the help of (9)
\bqry
K_0=\sum _{k=1}^N\frac{-E_k^2/c^2+{\bf p}_k^2}{2M_k} & = & \sum _{k=1}^N
\{-\frac{1}{2M_kc^2}\epsilon _k(\epsilon _k+2M_kc^2)-\frac{M_k
c^2}{2}+\frac{{\bf p}_k^2}{2M_k}\} \NL   & = &  \sum _{k=1}^N\frac
{{\bf p}_k^2}{2M_k}-\sum _{k=1}^N\epsilon _k-\frac
{c^2}{2}\sum _{k=1}^NM_k-\sum _{k=1}^N\frac{\epsilon _k^2}{2M_kc^2}.
\eqry
Since, according to the equations of motion, $$\frac{d}{d\tau }ct_k=-
\frac{\partial K_0}{\partial (E_k/c)}=-\frac{\partial K_0}{\partial
(\epsilon _k/c)}=c+\frac{\epsilon _k}{M_kc},$$ it
follows that $$ct_k=c\tau +\int _0^\tau \frac
{\epsilon _k(\tau ^{'})}{M_kc}d\tau ^{'}+ct_k(0);$$ choosing
now in the initial instant $t_k(0)=t(0)$ for all of the particles,
$k=1,\ldots ,N,$ and taking $c\rightarrow \infty ,$ we obtain (18).

Finally, taking into account all of the abovementioned considerations, we
can see that the initial wave function $\psi _\tau (q_1,\ldots ,q_N)$ in
the Galilean limit approaches the nonrelativistic expression
\beq
\psi _\tau ({\bf r}_1,\ldots ,{\bf r}_N,t)=(\frac{1}{2\pi \hbar })^
{3N}\int d^3{\bf p}_1\cdots d^3{\bf p}_Ne^{\frac{i}{\hbar }\sum _{k=1}^N
({\bf p}_k{\bf r}_k-\frac{{\bf p}^2_k}{2M_k}t)}
e^{-\frac{i}{\hbar }\varphi }
\psi _0({\bf p}_1,\ldots ,{\bf p}_N),
\eeq
up to an additional phase factor $e^{-\frac{i}{\hbar }\varphi },$ where
$$\varphi =\frac{Mc^2t}{2},\;\;\;M=\sum _{k=1}^NM_k.$$

\section{Galilean limit of the relativistic Maxwell-Boltzmann
distribution}
We now wish to consider the Galilean limit of the equilibrium
relativistic Maxwell-Boltzmann distribution used in ref. [1],
\beq
f_0(q,p)=C(q)e^{A(p+p_c)^2},
\eeq
which is normalized as follows,
\beq
\int d^4pf_0(q,p)=n(q),
\eeq
where $n(q)$ is the total number of events
per unit space-time volume in the neighborhood of the point $q.$

Since in the framework of the relativistic Boltzmann theory [1],[5]
particles are considered as having equal intrinsic parameters,
$$M_1=M_2=\ldots =M_N=M.$$ It then follows from the relations (we
suppress $c$ for the present consideration) $$\eta =m-M,\;\;\;
-\triangle \leq \eta \leq \triangle $$ that
\beq
M-\triangle \leq m\leq M+\triangle .
\eeq
Since $\triangle $ may take any infinitesimal value\footnote{It
corresponds to infinitely sharp mass shell $\triangle =\mid m-M\mid .$}
but not zero, i.e., the variation in mass of the particles of the
ensemble may be very small, we can take the value of $p^2\equiv -m^2$
restricted to a small neighborhood of a fixed value $-M^2.$ This permits
us to write (21) as
\beq
f_0(q,p)\cong C(q)e^{-A(M^2+m_c^2)}e^{2Ap^\mu p_{c\mu }}.
\eeq
Introducing hyperbolic variables [5] and performing integration
\cite{GrRy}, we obtain from (22) and (24) the normalization relation
\beq
n(q)=C(q)\frac{4\pi \triangle M^2}{Am_c}e^{-A(M^2+m_c^2)}K_1(2AMm_c),
\eeq
where $K_1$ is the Bessel function of the third kind (3).

Identifying Synge's Lagrange parameters [3],
\beq
\xi ^\mu =2Ap_c^\mu ,
\eeq
and hence $\xi =\sqrt{-\xi ^\mu \xi _\mu }=2Am_c,$ we see that (24)
coincides with the result of Synge [3] obtained directly from an
on-mass-shell relativistic kinetic theory.

The average value of $p^\mu $ can be obtained, similar to [1], using the
parametric differentiation with respect to $p_c^\mu :$
\beq
\langle p^\mu \rangle _q=p_c^\mu \frac{M}{m_c}\frac{K_2(2AMm_c)}
{K_1(2AMm_c)}.
\eeq
The absolute temperature is defined, similar to [1],[5], through the
relation
\beq
2Am_c=\frac{1}{k_BT},
\eeq
which implies that in thermal equilibrium $Am_c$ is independent of $q;$
so that
\beq
\langle p^\mu \rangle =p_c^\mu \frac{M}{m_c}\frac{K_2(M/k_MT)}{K_1(M/k_BT
)}.
\eeq
As in [1],[5], to
obtain the local energy density we make a Lorentz transformation to
the rest frame of the local average motion. According
to (29), the relative velocity of the new frame is $${\bf u}=\frac{{\bf p
_c}}{m_c}.$$ The rest frame energy is then $$\langle E^{'}\rangle =\frac
{\langle E\rangle -{\bf u}\cdot {\bf p}}{\sqrt{1-{\bf u}^2}},$$ so that
\beq
\langle E^{'}\rangle =M\frac{K_2(M/k_BT)}{K_1(M/k_BT)}.
\eeq
Using the asymptotic formula [11,(9.7.2)]
\beq
K_\nu (z)\sim \sqrt{\frac{\pi }{2z}}e^{-z}\{1+\frac{4\nu ^2-1}{8z}+\cdots
\},\;\;\;z\rightarrow \infty ,
\eeq
we obtain, for $T\rightarrow 0,$
\beq
\langle E^{'}\rangle -M\cong \frac{3}{2}k_BT,
\eeq
in agreement with the Dulong-Petit law.
On the other hand, for $T\rightarrow \infty ,$ one may use another
asymptotic formula [11,(9.6.9)]
\beq
K_\nu (z)\sim \frac{1}{2}\Gamma (\nu )\left(\frac{z}{2}\right)^{-\nu },
\;\;\;z\rightarrow 0,
\eeq
to obtain
\beq
\langle E^{'}\rangle \cong 2k_BT,
\eeq
the result previously obtained in [1],[5],[6].

In conclusion we shall show that in the Galilean limit the variable
$E-m=\frac{{\bf p}^2}{2M}$ has the usual nonrelativistic
Maxwell-Boltzmann distribution.

Since the normalization
conditions for the low-temperature and the sharp-mass forms of the
relativistic Maxwell-Boltzmann distribution (21)
read\footnote{It follows from (1) and (24), respectively.}
$$n(q)=C(q)e^{-Am_c^2}\int d^4pe^{2Ap^\mu p_{c\mu }}$$ and
$$n(q)=C(q)e^{-A(M^2+m_c^2)}\int d^4pe^{2Ap^\mu p_{c\mu }},$$ all the
difference between these two forms is contained in the normalization
factor. The remaining integral on $d^4p$ can be written in the local rest
frame $p_c^\mu =(m_c,{\bf 0})$ $(2Am_c=\frac{1}{k_BT}):$
$$\int dEd^3{\bf p}e^{-\frac{E}{k_BT}}.$$ Taking into account (12), one
can rewrite this expression as follows: $$\int d^3{\bf
p}dme^{-(m+\frac{{\bf p}^2}{2M})/k_BT}=\int d^3{\bf p}
e^{-\frac{{\bf p}^2}{2M}}\int dme^{-\frac{m}{k_BT}}.$$ The latter
integral $$\int_{M-\triangle }^{M+\triangle }dme^{-\frac{m}{k_BT}}=
2k_BTe^{-\frac{M}{k_BT}}\sinh \frac{\triangle }{k_BT}$$ does not
vanish since $\triangle $ is finite (it enters the normalization factor).
We see that the freedom of $\triangle $ to take any value (in this case
not necessary infinitesimal but $\leq M),$
finite as $c\rightarrow \infty $ but not equal to zero, enables
one to obtain the nonrelativistic Maxwell-Boltzmann distribution
for $e\equiv \frac{{\bf p}^2}{2M}.$

Finally, we have for the low-temperature form\footnote{For the sharp-mass
form there is an additional factor $e^{-AM^2}$ in the r.h.s. of (35).} of
the relativistic Maxwell-Boltzmann distribution (21),
\beq
n(q)=2C(q)k_BT\sinh
\frac{\triangle }{k_BT}e^{-\frac{m_c}{2k_BT}}e^{-\frac{M}{k_BT}}
\int d^3{\bf p}e^{-\frac{{\bf p}^2}{2M}},
\eeq
which is the usual
(normalized) nonrelativistic Maxwell-Boltzmann distribution
$$f(e)=\frac{1}{\Gamma (\frac{3}{2})
(k_BT)^{3/2}}e^{1/2}\exp (-\frac{e}{k_BT}).$$

\section{Concluding remarks}
We have considered the Galilean limit of equilibrium relativistic
ensemble.
We have found that the relativistic relation between the energy $E$ and
the mass $m$ transforms in this limit to $E=m+\frac{{\bf p}^2}{2M},$
giving rise to the nonrelativistic Maxwell-Boltzmann distribution of
$\frac{{\bf p}^2}{2M}.$ The first moment of this distribution
$\langle \frac{{\bf p}^2}{2M}\rangle $ (which coincides with
$\langle E-m\rangle )$ is equal to $\frac{3}{2}k_BT,$ in agreement with
the Dulong-Petit law, and no relativistic correction appears in this
limit, in contrast to [1].

For the case of an
equilibrium relativistic ensemble of $indistinguishable\;
events$ \cite{BH2} the distribution function is found to be
$$f(q,p)=C(q)\frac{1}{e^{-A(p+p_c)^2}\pm 1};$$ in the Galilean limit it
becomes the usual nonrelativistic distribution of Bose-Einstein or
Fermi-Dirac, with chemical potential $\mu _G=\mu -M,$ where $\mu $
is the chemical potential of relativistic theory [6] and $M$ is the
Galilean mass.


\begin{thebibliography}{9}
\bibitem{BH} L. Burakovsky and L.P. Horwitz, Physica A, {\it in press}
\bibitem{Hak} R. Hakim, J. Math. Phys.  {\bf 15} (1974) 1310
\bibitem{Syn} J.L. Synge, The Relativistic Gas,  (North-Holland,
Amsterdam, 1957)
\bibitem{LevyL} J.M. Levy-Leblond, J. Math. Phys.  {\bf 4} (1963) 776
\bibitem{HorShaSch} L.P. Horwitz, S. Shashoua and W.C. Schieve,
     Physica A  {\bf 161} (1989) 300
\bibitem{HSP} L.P. Horwitz, W.C. Schieve and C. Piron,
     Ann. Phys. (N.Y.)  {\bf 137} (1981) 306
\bibitem{HR} L.P. Horwitz and F.C. Rotbart, Phys. Rev. D {\bf 24} (1981)
2127
\bibitem{rqm} L.P. Horwitz and C. Piron, Helv. Phys. Acta
{\bf 46} (1973) 316
\bibitem{conc} R. Arshansky, L.P. Horwitz and Y. Lavie,
     Found. Phys.  {\bf 13} (1983) 1167
\bibitem{GrRy} I.S. Gradshteyn and I.M. Ryzhik, Tables of Integrals,
Series, and Products (Academic Press, New York, 1980) p.358, subsection
3.547, formula 9
\bibitem{AS} M. Abramowitz and I.A. Stegun, Handbook of Mathematical
Functions, p.375, (Dover, New-York, 1970)
\bibitem{BH2} L. Burakovsky and L.P. Horwitz, {\it in preparation}
\end{thebibliography}
\end{document}